# Éléments pour la conception de Jeux Éducatifs sur Mobile

Aous Karoui[1], Iza Marfisi-Schottman[1], Sébastien George[1]
[1] LUNAM Université, Université du Maine, EA 4023, LIUM, 72085 Le Mans, France
aous.karoui@univ-lemans.fr, iza.marfisi@univ-lemans.fr, sebastien.george@univ-lemans.fr

**Résumé.** Les établissements scolaires sont de plus en plus équipés de dispositifs mobiles. Pourtant, des études récentes ont montré que leur utilisation par les enseignants reste encore limitée à cause d'un manque de formation, de ressources et d'applications adaptées pour des utilisations pédagogiques. Parmi les nouvelles pistes d'utilisation, les recherches sur les Jeux Éducatifs sur Mobile (JEM) semblent très prometteuses. Dans cet article, nous proposons d'identifier les JEM qui sont issus des travaux de recherches les plus référencés et qui ont été expérimentés, afin d'étudier leurs propriétés et impacts. Ceci nous permet, dans un deuxième temps, de définir les caractéristiques communes de ces JEM et d'identifier les principaux problèmes d'utilisation par les enseignants.

**Mots-clés.** jeu éducatif, mobilité, scénarisation pédagogique, apprentissage situé, enseignant

**Abstract.** School classrooms are equipped with more and more mobile devices. However, the latest studies show that their use by teachers is still very limited because of a lack of resources and applications adapted to use these artifacts in education. Among the new directions of use, research conducted on Mobile Learning Games (MLGs) seems very promising. In this article we propose to identify the MLGs from the most referenced research articles and that have proven their value in educational context, in order to determine their features and impact on learners. This study will then allow us to define the common characteristics of these MLGs and identify the main problems that occur when they are used by teachers.

## 1 Introduction

Le développement des technologies mobiles telles que les tablettes et les smartphones attire progressivement l'attention des responsables de l'éducation sur le plan national. En novembre 2014, le Président français a annoncé dans une interview télévisée, que toutes les classes de 5ème seront équipées de tablettes tactiles à la rentrée de 2016. Cette déclaration fait probablement suite aux retours positifs des campagnes d'expérimentations sur les tablettes dans les établissements scolaires lancées depuis 2010 par le ministère de l'Education et les collectivités territoriales [1]. Sur le plan mondial, d'autres gouvernements ont également pris la même initiative. Au Québec (Canada), plus de 10000 élèves utilisent quotidiennement une tablette en classe d'après une étude réalisée par Karsenti & Fievez [2]. Aux Etats-Unis, une autre étude montre que 43% des enseignants et/ou leurs élèves utilisent des liseuses électroniques et des tablettes en classe [3].

Face à la prolifération des artefacts mobiles en milieu scolaire, de récents travaux de recherche ont étudié leurs utilisations réelles en classe et leurs impacts sur les élèves et les enseignants. Malgré un effet d'enthousiasme observé pour les activités numériques sur tablettes et une amélioration de la communication et du travail collaboratif entre les élèves et les enseignants [4], les enseignants réclament de l'aide au niveau de la création de scénarios et de ressources pédagogiques, et aussi concernant le suivi des élèves pour leur évaluation [5].

Parallèlement, le développement des Jeux Educatifs sur Mobile (JEM) est une piste qui suscite l'intérêt de nombreux chercheurs. En effet, l'omniprésence des jeux sur mobile (12,8 millions de personnes en France, jouent sur leurs appareils mobiles [6]) et leurs nouvelles facultés telles que les activités situées [7] et la collaboration [8], rendues possibles grâce aux nouvelles technologies, ont motivé de récents travaux sur l'intégration des jeux éducatifs sur les tablettes et les smartphones [9]. Toutefois, nous pensons qu'il existe encore plusieurs verrous lors de la mise en pratique de ces JEM et que des solutions doivent être trouvées pour pallier les problèmes rencontrés par les enseignants en termes de conception, de scénarisation et de suivi des apprenants.

Afin de répondre à ces questions, nous proposons, dans cet article, d'analyser cinq JEM, qui ont été évalués en pratique dans des contextes scolaires et qui sont issus des travaux de recherches les plus référencés, dans le but d'en extraire les caractéristiques communes et les éléments qui ont favorisé la motivation et l'amélioration de l'apprentissage lors des expérimentations. Nous identifions également les problèmes récurrents lors de leur utilisation par les enseignants. En conclusion, nous discutons d'un modèle générique qui permet à la fois de réunir ces caractéristiques communes et de remédier aux problèmes identifiés.

## 2    Recherche documentaire des Jeux Éducatifs sur Mobile

Dans cette partie, nous présentons tout d'abord notre méthode de recherche documentaire pour identifier des JEM ayant connu un certain succès. Nous décrivons ensuite la procédure de sélection des JEM étudiés.

### 2.1    Méthode de recherche documentaire

Afin d'identifier notre ensemble de présélection, nous avons choisi de cibler nos recherches uniquement sur les évaluations scientifiques de JEM, c'est-à-dire des jeux testés et expérimentés en situation pédagogique. Pour délimiter les résultats de l'enquête, nous nous sommes basés sur les mots clés suivants : "*evaluation*"+"*mobile*"+"*learning game*". Ensuite, afin de couvrir le spectre des terminologies utilisées dans le domaine, nous avons remplacé le terme "*mobile*" par les termes suivants: "*location-based*", "*pervasive*", "*contextual*".

Notre méthode de recherche consiste à présélectionner un ensemble de JEM parmi les premiers résultats obtenus suite à la saisie des mots clés décrits ci-dessus, sur quatre moteurs de recherche scientifiques majeurs: *IEEExplore, ACM Digital Library, Science Direct et Springer*. Afin d'obtenir des articles qui n'auraient pas été trouvés par ces moteurs, nous avons également choisi d'étendre nos recherches documentaires au méta-moteur *Google Scholar*.

A ce stade, nous avons choisi de retenir uniquement les articles qui traitent des JEM informatisés et évalués selon les heuristiques d'évaluation définis par Zaibon et Shiratuddin [10] en termes d'utilisabilité, de mobilité, d'aspect ludique et de contenu pédagogique. De plus, nous nous somme focalisé sur les JEM mettant en œuvre des attributs spécifiques à la mobilité, telles que les activités situées, réalisables grâce aux techniques de localisation (notamment le GPS), ou des interactions entre les joueurs, rendues possibles à travers les techniques de connectivité (wifi ou réseaux mobiles).

Étant donné le nombre élevé de résultats, nous avons retenu uniquement les dix premiers JEM pour chaque moteur, triés par pertinence. Le tri par pertinence est employé par défaut par les 5 moteurs utilisés. Il correspond à l'intégralité du texte de l'article par rapport aux mots clés saisis et tient compte du nom de l'auteur et de la publication dans laquelle l'article est paru.

Le Tableau 1 contient les 50 résultats (5 moteurs x 10 résultats), ce qui donne en fait 39 JEM trouvés car certains apparaissent sur plus d'un moteur (en gras dans le Tableau 1) présélectionnés sur un total de 250 articles parcourus sur les cinq moteurs de recherches utilisés. Nous avons cités les JEM une seule fois même s'ils apparaissaient dans plusieurs articles sur le même moteur de recherche.

**Tableau 1.** Les 10 premiers JEM trouvés sur les différents moteurs de recherche.[1]

| IEEExplore | ACM DL | Science Direct | Springer | Google Scholar |
|---|---|---|---|---|
| Skillville | On the Edge | Bauboss | **HeartRun** | **The MobileGame** |
| **Lecture Quiz** | Chinese-PP game | **HeartRun** | ToneWars | **Gaius' Day** |
| **Skattjakt** | **Parrot Game** | QuesTInSitu | **Power Agent** | **Frequency1550** |
| Bagamoyo Caravan | **Frogger&Floored** | Frequentie1550 | MobileMath | **Skattjakt** |
| FreshUp | **Power Agent** | EarlyBird | **Gaius' Day** | **Parrot Game** |
| Cardinal direction | Kurio | Furio's | Detective Alavi | **Frogger&Floored** |
| Tower of London | Power Explorer | MSGs | Preserving Famosa fortress | **Power agent** |
| The Amazing City | iFitQuest | Reenactment | Mindergie | **Lecture Quiz** |
| CatchBob! | **Gaius's Day** | BoomRoom | Language Learning Game | MuseumScrabble |
| **The MobileGame** | TimeWarp | EasyLexia | Nat. Palace Museum Adventure | Mentira |

### 2.1 Méthode de sélection

Parmi l'ensemble de présélection, nous avons choisi de restreindre notre analyse dans cet article aux cinq premiers JEM qui ont été les plus référencés. En effet, nous pensons que ce critère nous permettra de sélectionner d'une façon neutre les JEM à analyser, qui ont attiré le plus l'attention des chercheurs.

---

[1] La liste complète des références des articles des JEM est disponible sur: http://www.univ-lemans.fr/~akaroui/tableau_jem.html

Pour ceci, nous avons donc utilisé l'outil de tri par nombre de citations de chaque article disponible sur *Google Scholar*. Nous avons ensuite calculé le nombre de citations total pour chaque JEM qui est équivalent à la somme des citations des articles traitant d'un même JEM.

Le Tableau 2 recense les cinq premiers JEM selon notre critère de tri.[2]

**Tableau 2.** Classement par nombre de citations total des JEM présélectionnés.

| JEM | Article de référence | Année de parution | Nombre de citations | Nombre de citations total |
|---|---|---|---|---|
| *Frequency1550* | [11] | 2009 | 146 | 313 |
| | [12] | 2009 | 75 | |
| | [13] | 2011 | 69 | |
| | [14] | 2007 | 12 | |
| | [15] | 2007 | 11 | |
| *The MobileGame* | [16] | 2005 | 212 | 240 |
| | [16] | 2005 | 28 | |
| *Gaius' Day* | [17] | 2008 | 104 | 179 |
| | [18] | 2008 | 35 | |
| | [19] | 2009 | 18 | |
| | [20] | 2008 | 11 | |
| | [21] | 2012 | 10 | |
| *Power Agent* | [22] | 2009 | 49 | 88 |
| | [23] | 2007 | 39 | |
| *Skattjakt* | [24] | 2008 | 39 | 76 |
| | [25] | 2008 | 34 | |
| | [26] | 2009 | 3 | |

## 3   Description et analyse des JEM retenus

Nous analysons dans l'espace de cet article uniquement les cinq JEM du tableau 2. Cela nous semble suffisant pour dégager des caractéristiques même si le travail pourra être étendu par la suite. Pour *Frequency1550* et *The MobileGame,* nous nous sommes appuyés sur deux évaluations rapportées par les différents articles. Pour les autres JEM, une seule évaluation est décrite dans les articles. Notre méthode d'analyse consiste à identifier tout d'abord le but pédagogique du JEM et le scénario du jeu. Nous décrivons ensuite le contexte de l'évaluation et interprétons les résultats.

### 3.1   *Frequency1550*

**But pédagogique.** *Frequency1550* est un jeu éducatif conçu pour apprendre l'histoire de la ville d'Amsterdam à l'époque du moyen âge.

**Description du scénario.** Dans la vieille ville d'Amsterdam, les joueurs se répartissent en groupes de 4 ou 5 personnes. Ces groupes sont ensuite divisés en deux équipes : une équipe de terrain, qui va inspecter les lieux et une équipe directrice, qui restera en poste derrière un ordinateur dans le bâtiment central (Le Waag d'Amsterdam). Chaque équipe de terrain est équipée d'un téléphone mobile permettant de prendre des photos des lieux et de les envoyer à l'équipe directrice afin que celle-ci décide de la stratégie à adopter et en tenant compte de la nature de la mission, oriente l'équipe du terrain vers l'endroit de la solution. Les auteurs classent les rôles possibles dans le jeu selon trois types de profils : le profil récepteur, lors de l'introduction de l'histoire du lieu et la réception des indices vidéo et textuels. Le profil constructeur partiel du scénario réservé aux joueurs qui font partie des équipes directrices qui, selon les recherches sur internent, orientent l'équipe du terrain vers les endroits clés pour réaliser les missions. Le profil acteur fait référence aux membres des équipes du terrain qui sont sur le lieu, et qui informent leurs équipes directrices de leur avancement en les envoyant des images et des informations des scènes observées.

---

[2] Le tableau complet de l'ordre de citations des JEM est disponible sur http://www.univ-lemans.fr/~akaroui/tableau_jem.html

**Contexte de la 1ère évaluation.** Le jeu a été expérimenté sur un effectif de 232 élèves de 12 à 16 ans. Un groupe témoin formé de 226 élèves du même âge ayant assisté à un cours d'histoire formel, a servi d'exemple de comparaison[11].

**Résultats de la 1ère évaluation.** Les résultats ont montré que les joueurs de *Frequency1550* ont obtenu de meilleurs scores lors du test des connaissances que ceux qui ont eu des cours classiques sur le sujet. D'après les auteurs, les joueurs ont pu acquérir plus de connaissances grâce à leur présence directe sur les lieux historiques. Les différents profils des joueurs ont impliqué différentes méthodes d'acquisition des informations, que ce soit en recevant l'information en tant que spectateur, en la cherchant sur internet en tant que constructeur de scénario ou bien en la trouvant directement sur le lieu en tant qu'acteur. Les auteurs ont également remarqué que certains joueurs avaient été distraits par les évènements externes au jeu, par exemple survenant dans la rue, et ont parfois perdu la vision de la structure générale du scénario du jeu. Cependant, il semblerait que même si les élèves ont échoué aux missions, l'expérience située leur a permis d'acquérir les compétences pédagogiques d'une manière indirecte.

D'après les interviews des joueurs, il est possible que participer au jeu pendant un seul jour ne soit pas suffisant pour atteindre le niveau d'attraction maximal.

**Contexte de la 2ème évaluation.** L'évaluation a été réalisée sur un effectif de 216 élèves de 12 à 16 ans, issus de 10 collèges aux Pays-Bas[12].

**Résultats de la 2ème évaluation.** Les résultats ont montré que l'attention des étudiants a varié selon les trois types de profils dans le scénario. En tant que récepteurs, les étudiants ont porté plus d'attention au fonctionnement du jeu et aux technologies qu'au récit introductif de l'histoire. De même, davantage d'attention a été accordée aux informations accessibles par les supports mobiles qu'aux informations disponibles sur le lieu pour les équipes de terrain (profil acteur).

En tant que constructeurs partiels du scénario, les membres de l'équipe directrice étaient plus motivés que ceux des équipes du terrain grâce à la diversité de leurs tâches (recherches sur internet, orientation des équipes du terrain) et du sentiment de contrôle. Inversement, les équipes de terrain ont expérimenté un manque de contrôle et de vision du jeu. Finalement, la collaboration a été très importante entre les deux équipes.

### 3.2 *The MobileGame*

**But pédagogique.** C'est un jeu utilisé pour faciliter l'orientation dans une université et faire connaitre ses différents départements ainsi que son périmètre.

**Description du scénario.** Le jeu consiste à guider les participants le long d'un parcours, et à travers plusieurs missions, afin de découvrir des endroits dans l'université tels que la bibliothèque, la cafétéria ou un laboratoire. Des tâches contextuelles sont incluses dans les missions comme la recherche d'un certain livre dans la bibliothèque ou d'une certaine personne dans un département. Les joueurs peuvent jouer individuellement ou bien se répartir les tâches dans des groupes de deux, trois ou quatre personnes. Les missions du jeu sont accomplies à travers les terminaux mobiles.

**Contexte de la 1$^{ère}$ évaluation.** L'évaluation a été réalisée sur 22 étudiants de 19 à 25 ans à l'université de Koblenz en Allemagne et s'est déroulée en deux étapes : la première évalue l'utilisabilité du jeu et la 2$^{ème}$ évalue sa valeur éducative[16].

**Résultats de la 1$^{ère}$ évaluation.** En ce qui concerne l'utilisabilité du jeu, les résultats de la 1ère étape de cette évaluation ont conduit à identifier des problèmes d'affichage des informations et de réactivité du jeu et ont permis donc d'améliorer l'interface pour la 2$^{ème}$ expérience.

Parmi les résultats de la 2$^{ème}$ étape, cinq joueurs ont dit qu'ils ont juste apprécié le jeu alors que la majorité des joueurs (17 sur 22) ont affirmé qu'ils aimeraient y rejouer à tout moment. L'effet d'immersion et de distraction atteint les meilleurs scores suite aux activités *map-navigation* (s'orienter à travers la carte électronique) et *hunting and hiding* (éviter les groupes chasseurs et pourchasser les groupes proies visibles sur la carte électronique).

**Contexte de la 2$^{ème}$ évaluation.** L'évaluation a été réalisée sur 149 étudiants sur le campus universitaire de Zurich[27].

**Résultats de la 2$^{ème}$ évaluation.** Après analyse des résultats, les auteurs ont remarqués des différences significatives liées aux nombres de participants dans les équipes. En effet, les équipes composées de deux joueurs avaient les meilleurs résultats en apprentissage, *team-building* et réussite d'activités. Les équipes composées de quatre joueurs étaient en deuxième position avec une légère baisse d'apprentissage. Après une analyse plus détaillée du comportement des joueurs, il est aussi apparu que, dans les équipes de quatre, les joueurs se divisaient spontanément en deux groupes, ce qui peut expliquer leurs résultats positifs. Enfin, les équipes de trois avaient les scores les plus bas. Les auteurs expliquent ce résultat par le fait que, dans une équipe de trois, un des joueurs peut être moins actif à cause d'une mauvaise répartition des tâches.

### 3.3 *Gaius' Day (journée de Gaius)*

**But pédagogique.** Le jeu a été conçu dans le but d'améliorer l'apprentissage lors de la visite du parc archéologique d'*Egnazia* en sud d'Italie.

**Description du scénario.** Les joueurs se répartissent en groupes de 3 à 5 élèves, afin de jouer le rôle d'un citoyen romain appelé *Gaius*, qui s'installe à *Egnazia* avec sa famille. Le jeu existe en deux versions. La 1$^{ère}$ est une version papier où les joueurs sont équipés de brochures et de documents contenant la carte et la description des missions. La 2$^{ème}$ version est implémentée sur des téléphones mobiles à l'aide de l'outil auteur des jeux sur les sites archéologiques *Explore!* Cette version se caractérise par des effets sonores et des techniques de réalité augmentée exécutées sur les supports mobiles.

**Contexte de l'évaluation.** L'évaluation consiste à expérimenter le jeu *Gaius' Day* sur mobile à l'aide de l'outil auteur *Explore!* et comporte deux expériences. La 1ère a été réalisée sur un effectif de 24 élèves entre 11 et 13 ans, ayant déjà visité le même parc de façon formelle. La 2ème a été réalisée sur 42 élèves de 12 ans ayant la même expérience du premier groupe mais sur un autre site historique. Dans les deux expériences, les élèves devaient répondre à des questions pendant le jeu via leurs smartphones[17].

**Résultats de l'évaluation.** Les premiers résultats ont montré davantage de réussite dans les missions lors de la version papier. Les auteurs expliquent cette différence par le fait que la version *Explore!* n'affichait que les informations liées à la position GPS exacte des élèves alors qu'en version papier les élèves avaient l'ensemble des données à leur disposition. De plus, la version électronique obligeait les élèves à répondre selon l'ordre d'affichage des questions alors qu'en version papier, ils pouvaient choisir l'ordre de leurs réponses et revenir sur leurs réponses plus tard.

Concernant le niveau de la satisfaction des élèves, la motivation était très élevée dans les deux versions du jeu mais les facteurs d'attractivité évoqués par les joueurs étaient plus nombreux pour la version électronique. Ces facteurs concernaient souvent les artefacts utilisés (les effets sonores et de réalité augmentée) alors que sur la version papier, les facteurs d'attraction étaient liés au parc archéologique.

Le débriefing de la version électronique du jeu a duré 10 minutes de plus que le débriefing suivant la version papier. La reproduction en 3D des monuments historiques et l'exposition virtuelle des résultats de réponses de chaque groupe ont été très appréciés et ont permis d'échanger davantage.

Finalement, les deux versions ont été réussies au niveau de la motivation et de la collaboration entre les élèves. Par contre, la version électronique, bien qu'elle ait obtenu plus de commentaires positifs, ses limites cruciales au niveau de la présentation des informations et des questions a permis une dégradation dans la résolution des problèmes comparée à la version papier.

### 3.4 Power Agent

**But pédagogique.** L'objectif du jeu est de sensibiliser les joueurs aux énergies et de leur apprendre les bonnes habitudes de consommation.

**Description du scénario.** Les joueurs jouent le rôle d'un agent spécial qui a pour mission de faire des économies d'énergie dans son foyer. Pour atteindre ce but, les agents agissent sous la supervision de « Mr. Q », le directeur des agents. Chaque agent est équipé d'un téléphone mobile connecté directement au compteur électrique du foyer et doit coopérer avec les membres de sa famille pour baisser la consommation entre 17h et 22h. Ensuite, ses efforts sont combinés avec ceux d'autres agents dans la même ville. L'équipe d'agents spéciaux est en concurrence avec une autre équipe dans une autre ville. L'équipe gagnante est celle qui réalise la plus importante baisse de consommation d'énergie.

**Contexte d'évaluation.** Deux équipes de trois agents ainsi que les membres de leurs familles ont participé au jeu pendant 10 jours dans deux villes différentes en Suède pendant le printemps 2008. Par coïncidence, la 1ère équipe était composée uniquement de garçons et tous les membres de la 2èm équipe étaient des filles.[22]

**Résultats d'évaluation.** Les participants ont été très motivés et engagés dans le jeu et ont accepté un niveau de confort quotidien inférieur au standard. Selon les auteurs, la répartition des agents en une équipe de filles et une équipe de garçons a amélioré l'aspect compétitif qui a provoqué des initiatives remarquables chez certains agents comme le remplacement des lampes par des ampoules économiques ou bien l'utilisation des bougies dans certains cas. La collaboration familiale a aussi été un facteur immersif important ; les parents ont été très impliqués dans le jeu et communiquaient même avec « Mr. Q » à travers leurs enfants. L'objectif pédagogique a aussi été atteint dans la plupart des cas. Les joueurs ont appris de nouvelles méthodes pour réduire la consommation à travers les instructions suggérées par les échanges de commentaires sur la plateforme du jeu, par l'expérience située en appliquant les stratégies et aussi par les discussions avec les membres de famille et d'équipe. Bien qu'une baisse de consommation à long terme n'ait pas pu être démontrée par cette évaluation, le jeu a conduit des changements de comportement considérables pendant son déroulement.

### 3.5 *Skattjakt (chasse au trésor)*

**But pédagogique.** Le jeu a été conçu dans le but de promouvoir l'activité physique des jeunes tout en apprenant l'histoire d'un château situé sur le campus universitaire.

**Description du scénario.** Dans le scénario, les joueurs doivent aider une femme fantôme à résoudre le mystère de son mari disparu qui a construit le château. En explorant la carte s'affichant sur leurs smartphones, les joueurs doivent se rendre vers des points de repères afin de résoudre des puzzles, décrypter des codes de façon collaborative ou trouver d'autres indices les orientant vers de nouveaux endroits. Durant le jeu, les élèves reçoivent sur leurs appareils mobiles des indices textuels et sonores qui leur permettent d'avancer dans le jeu.

**Contexte d'évaluation.** L'étude à laquelle nous nous intéressons a été réalisée dans un premier temps sur 12 élèves entre 12 et 15 ans en février 2007. Dans un deuxième temps, 26 filles entre 13 et 15 ans on fait l'objet d'une seconde expérience avec le même jeu en juin 2007.[24]

**Résultats d'évaluation.** 58% des joueurs ont affirmé que le jeu était très divertissant lors de la 1$^{ère}$ expérience et 73% ont dit qu'il l'était lors de la deuxième. Ceci est aussi lié aux conditions du jeu (soirée hivernale pendant la 1$^{ère}$ expérience et une journée ensoleillée pendant la 2$^{ème}$). Les aspects divertissants ont aussi augmenté la motivation d'apprendre. En effet, 100% des joueurs ont été motivés pour apprendre l'histoire locale à travers le jeu lors de la 2$^{ème}$ expérience, 75% l'avaient été durant la 1$^{ère}$ expérience. Finalement, la collaboration a été également meilleure pendant le deuxième test. En effet, les résultats indiquent que 92% des joueurs (uniquement des filles) du second test ont collaboré ensemble tout le temps et 58% des joueurs ayant participé au premier test ont collaboré ensemble tout le temps.

## 4  Discussion

Tous les JEM ont obtenu des résultats positifs au niveau de la satisfaction des joueurs. Nous allons donc tout d'abord, identifier les éléments communs entre ces jeux qui ont contribué à cette réussite au niveau de l'expérience du jeu. Dans un deuxième temps, nous exposons les problèmes rencontrés par les joueurs lors des expériences afin de les analyser d'un point de vue enseignant.

**L'activité située et l'apprentissage.** C'est une activité commune à tous les jeux analysés. En effet, les différents scénaristes ont choisi l'expérience située comme moyen d'apprentissage. Nous avons des résultats d'apprentissage meilleurs que ceux à travers les cours formels dans le cas de *Frequency1550* notamment avec le type de profil acteur. De plus, il semble que les joueurs ont appris parfois d'une manière indirecte même s'ils n'étaient pas en train d'exécuter les missions du jeu, à travers leur relation directe (présence sur terrain) avec l'environnement d'apprentissage.

D'autres résultats concrets, dans le cas de *Power Agent*, où le but pédagogique a été atteint pour les deux équipes participantes. En effet, nous pensons que le fait de pouvoir pratiquer les méthodes acquises sur le lieu est plus efficace que de les acquérir uniquement d'une manière théorique.

Le but pédagogique a été aussi atteint d'après les auteurs pour *Gaius' Day* sans comparaison à l'enseignement formel puisque les deux expériences étaient situées.

Sous le thème de l'apprentissage situé, nous nous sommes intéressés aux propriétés intrinsèques des scénarios afin d'identifier les éléments les plus motivants. Ceci nous a permis d'exposer les éléments suivants.

<u>L'autonomie dans le jeu et la participation à la recherche de l'information:</u> Parmi les résultats des tests de *Frequency1550*, la motivation était plus importante chez les équipes directrices par rapport aux équipes de terrain à cause d'un degré d'autonomie dans le jeu au niveau de la recherche d'informations sur internet et de la possibilité de développer des stratégies par le fait d'orienter leurs co-équipiers. Les auteurs nous ont aidé à repérer cet aspect en identifiant cette activité comme étant « profil constructeur dans le jeu».

La liberté d'exploration a été concrètement évoquée comme un aspect motivant lors des interviews des joueurs ayant expérimenté *The MobileGame*. De plus, la liberté du choix de la stratégie pour l'ordre de réponses a permis d'avoir de meilleures réponses dans les résultats de *Gaius' Day* en version papier. Ceci nous laisse à priori recommander un degré d'autonomie pour les joueurs lors de la scénarisation, tout en prévoyant des moyens leur permettant de développer des stratégies et avoir un sentiment de contrôle. Nous recommandons aussi de privilégier l'autonomie de la recherche d'information plutôt que de la recevoir d'une manière statique. Ceci étant un atout de l'apprentissage situé par rapport à l'apprentissage théorique.

**La collaboration et la composition des équipes.** La mise en place d'activités collaboratives entre apprenants favorise l'engagement et l'immersion dans le jeu. Ceci a été un mécanisme présent dans tous les JEM analysés. En effet, tous les scénarios se sont basés sur la résolution de problème en groupe. Nous pensons que si cette caractéristique a été commune dans les jeux analysés, c'est que la collaboration est un moyen intéressant pour améliorer l'apprentissage à travers l'échange d'informations. En outre, l'implication de toute la famille dans *Power Agent* par exemple a favorisé l'immersion des joueurs. Les auteurs ont aussi reporté que la répartition hasardeuse des équipes en équipe de filles et équipe de garçons a amélioré la compétitivité. De plus, le travail collaboratif semble être plus réussi quand il s'agissait d'équipes composées uniquement de filles dans *Skattjakt,* qui ont collaboré d'une façon plus élevée que les équipes mixtes. Ceci nous laisse dire que dans certains cas, il est possible de tirer profit de la concurrence entre les sexes pour favoriser l'engagement. Par contre, il faut bien prendre en compte la tranche d'âge des joueurs et la nature des relations entre eux, c'est-à-dire leur niveau de connaissance.

D'autre part, le nombre de joueurs dans les équipes est aussi un élément à prendre en compte lors de la scénarisation. En effet, d'après la 2$^{ème}$ évaluation de *The*

*MobileGame*, les équipes en binôme ont réalisé les meilleurs scores dans le jeu parce qu'ils se sont bien répartis les tâches. Logiquement, une équipe composée de trop de joueurs mettra plus de temps dans la discussion et la répartition des tâches que des équipes de binômes. Par contre, nous estimons que cela impacte directement l'expérience sociale qui deviendra dans ce cas moins importante. Ainsi, nous recommandons d'ajuster l'effectif selon le temps accordé, l'objectif et la nature du scénario ; c'est-à-dire privilégier des équipes d'effectif réduit lorsque l'objectif est de réaliser des tâches dans un temps réduit et augmenter l'effectif dans les équipes si les conditions permettent plus de temps de réflexion et que le scénario n'est pas très dense. D'après l'expérimentation de *The MobileGame*, il semblerait qu'il soit préférable de privilégier un nombre pair d'apprenants pour composer les équipes. Toutefois, ceci reste à confirmer en étudiant plus d'expérimentations focalisées sur les compositions des équipes.

**L'effet attractif des nouvelles technologies.** La satisfaction est un élément important lors de l'expérience utilisateur et les nouvelles technologies y jouent un rôle principal. Nous avons noté que la motivation était toujours supérieure lorsqu'il s'agissait d'utiliser des outils techniques attractifs. En effet, les joueurs étaient plus attentifs et réactifs pendant l'explication du fonctionnement de leurs dispositifs que lors de la présentation de l'histoire du jeu (*Frequency1550*). Les interactions en réalité augmentée ont beaucoup motivé les joueurs de *Gaius' Day* dans la version *Explore !* et les représentations en 3D ont permis d'avoir des phases de débriefing et d'échanges plus riches que dans la version papier.

Par contre, les moyens techniques sont parfois un obstacle pour les JEM. Des problèmes d'utilisabilité d'interface ont été rapportés pour *Gaius' Day* (version *Explore!*) et *The MobileGame*.

**Les problèmes d'utilisation.** Dans cette partie, nous allons partir des problèmes rencontrés par les joueurs lors des expériences de jeu afin de les analyser d'un point de vue enseignant et proposer des pistes de solutions.

**Au niveau de l'exécution.** Pour l'exemple de *Frequency1550*, les joueurs qui pouvaient interagir dans le scénario en donnant des instructions aux autres ont éprouvé plus de satisfaction. Alors que les autres joueurs ayant moins de contrôle, ont parfois perdu les repères dans le jeu à cause d'un manque de concentration pendant l'introduction et quelques incidents survenus lors du jeu (pour les équipes de terrain). Nous estimons ici que le rôle de suivi d'avancement et d'intervention dans le scénario est en premier lieu, destiné à l'enseignant, afin qu'il puisse recadrer assez vite les joueurs en difficultés. Pour ceci, il est assez utile d'utiliser des outils de traçabilité qui permettent de suivre la progression des élèves dans le jeu et de prévoir des étapes alternatives dans le scénario pour gérer les différents niveaux de progression.

**Au niveau de la conception.** Les utilisateurs de *Skattjakt* ont regretté qu'il ne fût pas possible de jouer au même jeu pour d'autres thèmes que l'histoire. Ceci rejoint également des besoins exprimés par les enseignants, les auteurs de *Gaius' Day* (testé sur *Explore!*) en l'occurrence. Nous estimons que la généricité du scénario sera un caractère très avantageux dans le cas où l'outil auteur permettrait aux enseignants de créer leurs scénarios d'une façon plus générique. Des travaux de recherche se sont déjà engagés dans cette voie et ont permis aux enseignants de créer leurs propres scénarios à travers des méthodes simplifiées [28]. Néanmoins, ils ont été confrontés à des problèmes d'exécution notamment avec les couts de développement et les difficultés de coordination entre enseignants et développeurs.

## 5   Conclusion

Dans cet article, nous avons cherché à identifier les mécanismes de JEM qui ont contribué à leur succès, à travers l'analyse des évaluations de cinq JEM issus des travaux de recherches les plus cités. Ceci nous a permis en même temps de retrouver des problèmes récurrents d'utilisation, auxquels nous avons proposé des pistes de solutions.

Bien qu'elle nous ait permis d'identifier des JEM intéressants, notre méthode de recherche peut nous avoir fait passer à côté de JEM récents dont les articles n'ont pour le moment pas été beaucoup cités, mais qui rejoignent potentiellement plusieurs des mécanismes retrouvés dans la dernière partie de cet article. Par exemple, *Power Explorer* est un JEM qui améliore les mécanismes de *Power Agent* en enrichissant le processus d'apprentissage afin d'atteindre des changements d'habitudes de consommation à long terme.

La prochaine étape de notre travail consiste à définir un environnement auteur générique pour que des concepteurs non informaticiens (des enseignants, des conservateurs de musées, …) puissent scénariser leurs propres jeux et les déployer sur les dispositifs mobiles. Nous proposerons pour cela une bibliothèque d'activités types reprenant les caractéristiques identifiées dans cet article.

## Références